\documentclass[12pt,aps,prd,nofootinbib,noshowkeys,noshowpacs,longbibliography,
superscriptaddress,tightenlines,floatfix,preprintnumbers]{revtex4-2}
\usepackage{amsmath,amsfonts,amsthm,amssymb}
\usepackage{graphics,graphicx}
\usepackage{xcolor}
\usepackage{bm}
\definecolor{darkblue}{RGB}{0,0,196}
\definecolor{darkgreen}{RGB}{0,120,0}

\newcommand{\p}{\partial}

\def\beq{\begin{eqnarray}}
\def\eeq{\end{eqnarray}}

\usepackage[colorlinks=true,linktocpage=true,linkcolor=darkblue,citecolor=blue,urlcolor=darkblue]{hyperref}
\begin{document}
\preprint{CERN-TH-2025-121}

    \title{The perfect spinfluid: A divergence-type approach}
    \author{Nick Abboud}
    \email{nka2@illinois.edu}
    \affiliation{Illinois Center for Advanced Studies of the Universe \& Department of Physics, University of Illinois Urbana-Champaign, Urbana, IL 61801, USA}
    \author{Lorenzo Gavassino}
    \email{lorenzo.gavassino@gmail.com}
    \affiliation{Department of Mathematics, Vanderbilt University, Nashville, TN 37211, USA}
    \author{Rajeev Singh}
    \email{rajeev.singh@e-uvt.ro}
    \affiliation{Department of Physics, West University of Timisoara, Bulevardul Vasile P\^arvan 4, Timisoara 300223, Romania}
    \author{Enrico Speranza}
    \email{enrico.speranza@cern.ch}
    \affiliation{Department of Physics and Astronomy, University of Florence, Via G. Sansone 1, 50019 Sesto Fiorentino, Italy}
    \affiliation{Theoretical Physics Department, CERN, 1211 Geneva 23, Switzerland}

	\date{\today} 
	\bigskip

\begin{abstract}
We present a new formulation of non-dissipative relativistic spin hydrodynamics that incorporates spin degrees of freedom into the divergence-type theory framework. 
Due to the divergence-type structure, it is straightforward to enforce non-linear causality and symmetric hyperbolicity of the equations of motion, ensuring local well-posedness of the initial-value problem and stability of the theory.
Furthermore, in a specific realization based on spin kinetic theory, we prove that the equations of motion remain non-linearly causal and symmetric-hyperbolic to all orders in the spin potential, provided a specific thermodynamic constraint is satisfied.
This framework can be applied for numerical simulations to study the dynamics of spin-polarized fluids, such as the quark-gluon plasma in heavy-ion collisions.
\end{abstract}

\maketitle
\newpage
\tableofcontents
\section{Introduction}
\label{sec:introduction}

Divergence-type theories \cite{Liu1986,GerochLindblom1990} are a universal language for describing systems of quasi-conservation laws \cite{Peralta-Ramos:2009srp,Peralta-Ramos:2010qdp,GavassinoNonHydro2022,Baggioli:2023tlc} at zeroth order in the gradient expansion. The main idea is the following: Consider a collection of quasi-conserved four-currents $N^{\lambda A}$, where $\lambda$ is an ordinary spacetime index, and $A$ is a label counting the currents, and suppose that the fluid is in local quasi-equilibrium, so that the fluxes $N^{jA}$ are functions of $N^{0B}$. Then, there exists a unique \emph{thermodynamically consistent} way of constructing local constitutive relations $N^{jA}=N^{jA}(N^{0B})$ that does not involve derivatives~\cite{GerochLindblom1990}. The resulting system of balance laws $\partial_t N^{0A}+\partial_j N^{jA}(N^{0B})=R(N^{0A})$ is what we call a ``divergence-type theory''. Under some mathematical conditions that are straightforward to enforce, theories of this kind possess the following convenient features:
\begin{itemize}
\item Their equations of motion are non-linearly causal, and they can be recast in a symmetric hyperbolic form (which guarantees local well-posedness) \cite{GerochLindblom1990,Geroch_Lindblom_1991_causal}.
\item Equilibrium states are stable under small perturbations, both hydrodynamically and thermodynamically \cite{GavassinoGibbs2021,GavassinoCausality2021}, and the transport coefficients automatically respect Onsager symmetry conditions \cite{GavassinoCasmir2022}.
\item The theory admits a well-defined stochastic generalization in terms of an effective action, where random fluctuations fulfill the Crooks fluctuation theorem \cite{Mullins:2025vqa}.
\end{itemize}
In the light of the above convenient features, one may wonder whether it is possible to include spin degrees of freedom into divergence-type hydrodynamics. Indeed, the answer is affirmative, as we shall show here. By generalizing the generating function to spin-phase-space, we introduce spin potential (a Lagrange multiplier corresponding to spin angular momentum) which quantifies the spin degrees of freedom.
This divergence-type formulation of spin hydrodynamics offers a potential framework for understanding spin dynamics in relativistic fluids across the entire range of collision energies, with particular relevance at low energies~\cite{STAR:2021beb,STAR:2023nvo,STAR:2023eck}.

Recent measurements of particle spin polarization in relativistic heavy-ion collisions~\cite{STAR:2017ckg,STAR:2019erd,ALICE:2019onw,ALICE:2019aid,ALICE:2021pzu,STAR:2022fan} have prompted intense investigation into the formulation of relativistic hydrodynamics with spin, where spin degrees of freedom are included as additional hydrodynamic variables. This is done by requiring that fluid dynamics is not only dictated by the conservation of the energy-momentum tensor, as in standard hydrodynamics, but also by the conservation of the total angular momentum tensor~\cite{Jackiw:2004nm,Becattini:2011zz,Karabali:2014vla,Florkowski:2017ruc,Florkowski:2017dyn,Montenegro:2017rbu,Florkowski:2018fap,Hattori:2019lfp,Montenegro:2018bcf,Florkowski:2019qdp,Montenegro:2020paq,Gallegos:2020otk,Singh:2020rht,Bhadury:2020puc,Weickgenannt:2020aaf,Fukushima:2020ucl,Gallegos:2021bzp,Li:2020eon,Hu:2021pwh,Wang:2021ngp,Hongo:2021ona,Florkowski:2021wvk,Weickgenannt:2022zxs,Cao:2022aku,Cartwright:2021qpp,Gallegos:2022jow,Biswas:2023qsw,Weickgenannt:2023bss,Becattini:2023ouz,Ren:2024pur,Drogosz:2024gzv,Wagner:2024fhf,Chiarini:2024cuv,Wagner:2024fry,Singh:2024cub,Giacalone:2025bgm,Bhadury:2025fil,Sapna:2025yss,Singh:2025hnb, Ivanov:2025izv}.


In this work, we incorporate spin degrees of freedom into the divergence-type theory, extending its applicability to spin-polarized systems. {\color{black}We work with a specific expression of the generalized equation of state (generating function) for a kinetic theory of weakly coupled massive spin-$\frac{1}{2}$ particles by extending the Maxwell--Jüttner distribution function to the spinfluid and obtain the currents for particle number, energy-momentum, and spin.} We prove that the equations of motion are non-linearly causal and symmetric hyperbolic, making them suitable for numerical applications. A key feature of our approach is that non-linearly causal and symmetric hyperbolic properties hold to all orders in spin potential, provided the spin potential satisfies the constraint given in Eq.~\eqref{convergence}, which appears to be consistent with experimental values.

The structure of the article is as follows: we lay out the foundations of our theory in Sec.~\ref{sec:construction}, where we specify the need for divergence-type theory, mentioning its advantages such as symmetric hyperbolicity and causality. Then, in Sec.~\ref{sec:thermodynamicanalysis}, we study the thermodynamic properties of our formalism, followed by the explicit details of our framework in Sec.~\ref{sec:Generalized equation of state}. In Sec.~\ref{sec:Generalized equation of state} we also carry out a linearized analysis to demonstrate the causality and stability to all orders in spin potential. We check that our formalism reproduces the previously known results when truncated to first-order in spin potential. Finally, conclusions are given in Sec.~\ref{sec:conclude}.

\subsection{Notation and conventions}

We adopt the signature $(-,+,+,+)$ and work in natural units $c=\hbar=k_B=1$. Our spacetime is Minkowski, and we adopt standard rectangular coordinates, $x^\mu$ (spacetime indices are Greek). The labels $A,B,C$ are field multi-indices, defined in the text, while the labels $I,J$ are charge multi-indices (also defined in the text). Einstein's convention applies to all types of indices. 
We denote anti-symmetrization as $A^{[\mu\nu]}=\frac{1}{2}\left(A^{\mu\nu}-A^{\nu\mu}\right)$.
To avoid confusion, below we summarize our conventions:
\begin{itemize}
\item For all types of angular momentum, $\mathcal{J}^{\mu \nu}$ (total), $\mathcal{L}^{\mu \nu}$ (orbital), and $\mathcal{S}^{\mu \nu}$ (spin), we adopt the same convention as \cite{MTW_book,Speranza:2020ilk}: for a point particle with position $x^\mu$ and momentum $p^\mu$, we have that $\mathcal{L}^{\mu \nu}=x^\mu p^\nu-x^\nu p^\mu$.
\item For the thermal vorticity $\varpi_{\mu \nu}$, we fix the sign in such a way that $\varpi_{12}>0$ corresponds to a flow of particles with $\mathcal{L}^{12}>0$. This is equivalent to imposing $\varpi_{\mu \nu}=\partial_{[\mu} \beta_{\nu]}$. In global equilibrium, $\beta_\mu(x) =b_\mu +x^\nu \varpi_{\nu \mu}$.
\item For the spin potential $\Omega_{\mu \nu}$, we fix the sign in a way to resemble a standard chemical potential (in units of temperature), e.g. $dS = -\Omega_{12} \, d\mathcal{S}^{12}$, where $S$ is the entropy. Note that, to distinguish the spin from the entropy, we use a ``curly'' $\mathcal{S}$ for the former (e.g., the spin current is $\mathcal{S}^{\lambda \mu \nu}$). 
\end{itemize}

\section{Construction of the theory}
\label{sec:construction}

In this section, we lay out the foundations of the hydrodynamic theory for perfect spin-fluids.

\subsection{Basic assumptions and choice of pseudogauge}\label{basicassumptions}
For a general system of massive particles with spin-$1/2$, the total angular momentum $(\mathcal{J}^{\lambda\mu\nu})$ is decomposed in terms of orbital angular momentum $(\mathcal{L}^{\lambda\mu\nu})$ and spin angular momentum $(\mathcal{S}^{\lambda\mu\nu})$ as
\begin{eqnarray}
    \mathcal{J}^{\lambda\mu\nu} &=& \mathcal{L}^{\lambda\mu\nu} + \mathcal{S}^{\lambda\mu\nu}\,,\nonumber\\
    &=& x^\mu T^{\lambda\nu} - x^\nu T^{\lambda\mu} + \mathcal{S}^{\lambda\mu\nu}\,,
    \label{eq:totalAM}
\end{eqnarray}
where orbital angular momentum is decomposed in terms of energy-momentum tensor $\left(T^{\mu\nu}\right)$ which can have both symmetric and antisymmetric parts. Due to Noether's theorem, we have two conservation laws: the conservation of the energy-momentum tensor and the conservation of total angular momentum. This indicates that the spin angular momentum is not independently conserved, and we obtain the constitutive relation where the divergence of spin,
\begin{eqnarray}
    \partial_\lambda \mathcal{S}^{\lambda\mu\nu} &=& T^{\nu\mu} - T^{\mu\nu}\,,
    \label{eq:divergence_spin}
\end{eqnarray}
equals the antisymmetric parts of the energy-momentum tensor. However, during certain hydrodynamic processes, it is viable to assume, for an ideal fluid, two conditions:
\begin{itemize}
    \item[(i)] Entropy is conserved, i.e., the evolution is reversible.
    \item[(ii)] Spin-orbit coupling is of higher order in gradients and can be neglected if we remain in the ideal limit.
\end{itemize}
These conditions render the spin angular momentum independently conserved due to the vanishing antisymmetric parts of the energy-momentum tensor. We call this type of fluid a \emph{perfect spinfluid}.
Hence, we have two conservation laws at hand:
\begin{eqnarray}
    \partial_\mu T^{\mu\nu} = 0\,, \quad \partial_\lambda \mathcal{S}^{\lambda\mu\nu} = 0\,,
    \label{eq:conservation_laws}
\end{eqnarray}
where $T^{\mu\nu}$ is symmetric and $\mathcal{S}^{\lambda\mu\nu}$ is antisymmetric in its last two indices.\footnote{Note that, using Noether's theorem for a free Dirac Lagrangian, we obtain an asymmetric (canonical) energy-momentum tensor and a completely antisymmetric  (canonical) spin tensor. Therefore, Eq.~\eqref{eq:conservation_laws} forces us to choose a specific pseudogauge~\cite{Hehl:1976vr,Speranza:2020ilk,Singh:2024qvg} that makes the new energy-momentum tensor symmetric and the spin tensor antisymmetric only in the last two indices.}

An example of a physical system for which Eq.~\eqref{eq:conservation_laws} is valid is a gas of massive spin-$1/2$ fermions, where non-local collisions can be neglected. In this limit, within the framework of quantum kinetic theory, local collisions dominate, and as a result, there is no conversion between orbital and spin angular momentum. This leads to the conservation of the spin tensor, where conditions (i) and (ii) are fulfilled \cite{Florkowski:2018fap,Weickgenannt:2020aaf,Speranza:2020ilk,Das:2022azr}.

In perfect spinfluids, the energy-momentum tensor $T^{\lambda \mu}$ is symmetric. The proof is straightforward: the orbital angular momentum across a Cauchy hypersurface $\Sigma$ is, by definition,
\begin{equation}
    \mathcal{L}^{\mu \nu}(\Sigma)= \int_\Sigma \left(x^\mu T^{\lambda \nu}-x^\nu T^{\lambda \mu}\right) d\Sigma_\lambda \, ,
\end{equation}
where $d\Sigma_\lambda$ is the normal surface element (with standard orientation $d\Sigma_0>0$). Applying the Gauss theorem in the four-volume $\mathcal{V}$ between two Cauchy surfaces $\Sigma_1$ and $\Sigma_2$, we obtain~\cite{MTW_book}
\begin{equation}
    \mathcal{L}^{\mu \nu}(\Sigma_1)-\mathcal{L}^{\mu \nu}(\Sigma_2) = \int_{\mathcal{V}} \left(x^\mu \partial_\lambda T^{\lambda \nu}-x^\nu \partial_\lambda T^{\lambda \mu}+ T^{[\mu \nu]}\right) d^4 x \, .
\end{equation}
Imposing that $\mathcal{L}^{\mu \nu}$ does not depend on $\Sigma$, i.e. $\mathcal{L}^{\mu \nu}(\Sigma_1)=\mathcal{L}^{\mu \nu}(\Sigma_2)$ for all choices of $\Sigma_1$ and $\Sigma_2$, and invoking the energy-momentum conservation ($\partial_\lambda T^{\lambda \mu}=0$), we find that $T^{[\mu \nu]}=0$, which is what we wanted to prove.
\subsection{Necessity of a divergence-type framework}\label{framework}
Our goal, now, is to show that a hydrodynamic theory that is compatible with conditions (i) and (ii) should be of divergence type. Our argument is an adaptation of the original proof of~\cite{GerochLindblom1990} to the framework of spin hydrodynamics. Note, however, that the original idea can be traced back to~\cite{Friedrichs1974,Liu1986}.

In a non-dissipative theory, the relevant dynamical equations are the set of \emph{independent} conservation laws, which for perfect spinfluids are 
\begin{itemize}
    \item the particle number conservation, $\partial_\lambda N^\lambda=0$ ($N^\lambda$ is the particle-minus-antiparticle current),
    \item the energy-momentum conservation, $\partial_\lambda T^{\lambda \mu}=0$,
    \item and the spin conservation, $\partial_\lambda \mathcal{S}^{\lambda \mu \nu}=0$ ($\mathcal{S}^{\lambda \mu \nu}=\mathcal{S}^{\lambda [\mu \nu]}$ is the spin current).
\end{itemize}
The last one follows from condition (ii)\footnote{The conservation of orbital angular momentum is redundant, as it follows from the energy-momentum conservation, and from the symmetry of the stress-energy tensor, see Sec. \ref{basicassumptions}.}. These conservation laws can be collected together into one expression:
\begin{equation}\label{conservations}
    \partial_\lambda N^{\lambda A}=0 \, ,
\end{equation}
where $N^{\lambda A}= \{ N^\lambda, T^{\lambda \mu},\mathcal{S}^{\lambda \mu \nu} \}$ are the ``conserved fluxes'' of the theory, and $A$ is a multi-index. Since the theory is non-dissipative, the principles of relativistic thermodynamics~\cite{Israel_Stewart_1979,Callen_book,GavassinoTermometri} tell us that the entropy current $s^\lambda$ must be a function of the conserved fluxes $N^{\lambda A}$, and it must obey Israel's covariant Gibbs relation \cite{Israel_2009_inbook},
\begin{equation}\label{covariantGibbs}
    ds^\lambda = -\zeta_A \, dN^{\lambda A} \, ,
\end{equation}
for any variation $dN^{\lambda A}$.\footnote{The zeroth component of \eqref{covariantGibbs}, i.e. $ds^0=-\zeta_A dN^{0A}$, is the macroscopic limit of Zubarev's local equilibrium assumption \cite{Becattini:2014yxa}, according to which the system maximizes the entropy density $s^0$ for fixed values of the conserved densities $dN^{0A}$ at every spacetime event. The requirement that such a condition should hold in all reference frames gives the other components of \eqref{covariantGibbs}.} This immediately guarantees the reversibility of the dynamics: $\partial_\lambda s^\lambda = -\zeta_A \partial_\lambda N^{\lambda A}=0$. The fields $\zeta_A=\{\alpha, \beta_\mu, \Omega_{\mu \nu}/2 \}$ may be interpreted, respectively, as the fugacity, the inverse-temperature four-vector, and half of the spin potential (note that $\Omega_{\mu \nu}=-\Omega_{\nu \mu}$).

Let us now introduce the ``Massieu current'' $\chi^\lambda=s^\lambda+\zeta_A N^{\lambda A}$, i.e., the Legendre transform of the entropy current with respect to $\zeta_A$. Its differential reads
\begin{equation}\label{diffuzzo}
    d\chi^\lambda = N^{\lambda A} d\zeta_A \, .
\end{equation}
The key observation now is the following: Equation \eqref{conservations} constitutes a system of $1+4+6=11$ independent equations. 
On the other hand, $\zeta_A$ also has $1+4+6=11$ independent components. Therefore, we can treat $\zeta_A$ as the ``fundamental fields'' (i.e., the degrees of freedom) of the theory. This implies that, if we are given a formula $\chi^\lambda(\zeta_A)$, which expresses the Massieu current as a function of the fields, we can use the differential \eqref{diffuzzo} to derive the constitutive relations $N^{\lambda A}(\zeta_B)$. Their differential can be expressed in the form $dN^{\lambda A}=M^{\lambda A B}d\zeta_B$, and Eq. \eqref{conservations} becomes a system of 11 independent equations in 11 independent variables:
\begin{equation}\label{symmetrichyperb}
    M^{\lambda A B}\partial_\lambda \zeta_B=0 \, ,
\end{equation}
where $M^{\lambda AB}$ are algebraic functions of $\zeta_C$. Taking the exterior derivative  of Eq.~\eqref{diffuzzo}, we get $M^{\lambda[AB]}d\zeta_B \wedge d\zeta_A = 0$, which implies that $M^{\lambda AB}$ are symmetric under the exchange between $A$ and $B$. As a consequence, the system \eqref{symmetrichyperb} is symmetric. In Sec. \ref{symmuzzo}, we show that the system \eqref{symmetrichyperb} can be recast in a symmetric hyperbolic form, and it is causal \cite{Geroch_Lindblom_1991_causal} if and only if the vector 
\begin{equation}\label{causality}
    M^{\lambda AB}Z_A Z_B \,,
\end{equation}
is future directed and non-spacelike for any non-zero $Z_A=\{ Z,Z_\mu,Z_{[\mu \nu]} \}$. Therefore, the initial value problem is locally well-posed and the dynamics of the spinfluid can be solved for generic (regular) initial data on Cauchy surfaces~\cite{Kato2009}. 

\subsection{Symmetric hyperbolicity and causality}\label{symmuzzo}

To prove symmetric hyperbolicity, we need first to address a technical complication: the system \eqref{symmetrichyperb} is written in a symmetric form, but the variables $\zeta_A$ are not all independent. In fact, the system \eqref{symmetrichyperb} contains $1+4+16=21$ equations. However, the antisymmetry constraint on the spin potential makes $10$ equations redundant. Unfortunately, we cannot just ``throw away'' the redundant equations, because we may lose the manifest symmetry of \eqref{symmetrichyperb}. Luckily, there is an elegant solution. Working in some fixed coordinate system, we introduce $11$ unconstrained (i.e. independent) fields $\varphi^a,a=0,...,10$, which constitute the actual algebraic degrees of freedom of the theory. We treat these as our fundamental variables and we express the fields $\zeta_A$ as functions of $\varphi^a$ through some linear constitutive relations $\zeta_A(\varphi^a)$. For example, we can take $\alpha=\varphi^0$, $\beta_\mu=(\varphi^1,\varphi^2,\varphi^3,\varphi^4)$, and
\begin{equation}
    \Omega_{\mu \nu} = 
    \begin{bmatrix}
    0 & \varphi^5 & \varphi^6 & \varphi^7 \\
    -\varphi^5 & 0 & \varphi^8 & \varphi^9 \\
    -\varphi^6 & -\varphi^8 & 0 & \varphi^{10} \\
    -\varphi^7 & -\varphi^9 & -\varphi^{10} & 0 \\
    \end{bmatrix} \, .
\end{equation}
In this way, the constraint $\Omega_{(\mu \nu)}=0$ is automatically enforced at the level of the constitutive relations, and there is no constraint on  $\varphi^a$. Note that, since the constitutive relation $\zeta_A(\varphi^a)$ is linear, we can also express it in the form
\begin{equation}\label{smaug}
    \zeta_A = \dfrac{\partial \zeta_A}{\partial \varphi^a} \, \varphi^a \, ,
\end{equation}
where $\partial \zeta_A /\partial \varphi^a$ is a constant matrix. Let us now contract Eq.~\eqref{symmetrichyperb} with $\partial \zeta_A /\partial \varphi^a$, and use \eqref{smaug} to move from the variables $\zeta_B$  to the variables $\varphi^b$. The result is
\begin{equation}
    M^{\lambda A B} \dfrac{\partial \zeta_A}{\partial \varphi^a} \dfrac{\partial \zeta_B}{\partial \varphi^b}\partial_\lambda \varphi^b=0 \, .
\end{equation}
This is a system of 11 equations (since $a=0,...,10$) in the variables $\varphi^b$, which are, in turn, $11$ in number. It is also symmetric, due to the symmetry of $M^{\lambda AB}$. Symmetric hyperbolicity and causality require that \cite{Geroch_Lindblom_1991_causal}
\begin{equation}
    M^{\lambda A B} \dfrac{\partial \zeta_A}{\partial \varphi^a} \dfrac{\partial \zeta_B}{\partial \varphi^b} Z^a Z^b \,,
\end{equation}
be timelike future directed for any non-vanishing $Z^a=(Z^1,...,Z^{11})$. But it is immediate to verify that, if $Z^a$ is non-vanishing, then 
\begin{equation}
    Z_A= \dfrac{\partial \zeta_A}{\partial \varphi^a} Z^a \, ,
\end{equation}
is also non-vanishing. Hence, we only need to require that the vector introduced in Eq.~\eqref{causality} be future directed and timelike for any non-vanishing $Z_A=\{Z,Z_\mu,Z_{[\mu \nu]}\}$. This completes our proof.

We can make one last observation. If we fix the values of $\alpha$ and $\Omega_{\mu \nu}$, Eq.~\eqref{diffuzzo} becomes $d\chi^\lambda=T^{\lambda \mu}d\beta_\mu$. Taking the wedge product with $d\beta_\lambda$, we obtain $d\chi^\lambda \wedge d\beta_\lambda =0$ (recall that $T^{[\lambda \mu]}=0$). This implies that the one-form $\chi^\lambda d\beta_\lambda$ is exact, namely, there is a scalar field $\chi(\zeta_A)$ such that
\begin{equation}
    \chi^\lambda = \dfrac{\partial \chi}{\partial \beta_\lambda} \, .
\end{equation}
The scalar $\chi$ is known as the \emph{generating function} of the theory~\cite{GerochLindblom1990}.
Once it is assigned, all other quantities can be computed from it. In summary, the hydrodynamic theory that describes perfect spinfluids is a divergence-type theory \cite{Liu1986}, where the third flux,  $\mathcal{S}^{\lambda \mu \nu}$, is antisymmetric in $\mu$ and $\nu$ (instead of being symmetric traceless), and the energy-momentum tensor is symmetric, by construction.
\subsection{Overview on the equations of the theory}
\label{subsec:overview_EOM}
The construction of the theory outlined above is quite abstract, as it makes heavy use of the multi-index $A$. Here, we list the most important equations of the theory, expanding all the expressions that involve $A$.

The ``equation of state'' of the spinfluid is
\begin{equation}
   \chi=\chi(\alpha,\beta_\mu,\Omega_{\mu \nu}/2) \, ,
\end{equation}
which expresses the generating function in terms of the fundamental hydrodynamic fields of the theory. Its relation with statistical mechanical computations will be outlined in Sec.~\ref{statemech}. The conserved currents of the theory are computed as follows:
\begin{equation}\label{agmuni}
    \begin{split}
        N^\lambda ={}& \dfrac{\partial^2 \chi}{\partial \beta_\lambda \partial \alpha} \, , \\
        T^{\lambda \mu} ={}& \dfrac{\partial^2 \chi}{\partial \beta_\lambda \partial \beta_\mu} \, , \\
        \mathcal{S}^{\lambda \mu \nu}={}& \dfrac{\partial^2 \chi}{\partial \beta_\lambda \partial (\Omega_{\mu \nu}/2)} \, , \\
        s^\lambda ={}& \dfrac{\partial \chi}{\partial \beta_\lambda} -\alpha N^\lambda -\beta_\mu T^{\lambda \mu} -\dfrac{\Omega_{\mu \nu}\mathcal{S}^{\lambda \mu \nu}}{2} \, . \\
    \end{split}
\end{equation}
Recall that all the currents above have vanishing divergence.
Note that, in the computation of $\mathcal{S}^{\lambda \mu \nu}$, one needs to take extra care in computing the derivative with respect to $\Omega_{\mu \nu}$, because its components are not all independent. Derivatives with respect to  components of antisymmetric tensors are rigorously defined in appendix \ref{AAA}. The covariant Gibbs relation~\eqref{covariantGibbs} can be expanded as
\begin{equation}\label{covGibbsFinal}
    ds^\lambda =-\alpha dN^\lambda -\beta_\mu dT^{\lambda \mu} - \dfrac{\Omega_{\mu \nu} d\mathcal{S}^{\lambda \mu \nu}}{2} \, .
\end{equation}
The first two terms on the right-hand side are present also in perfect fluids without spin~\cite{Israel_Stewart_1979}, while the last term is specific to perfect spinfluids~\cite{Florkowski:2017ruc,Hattori:2019lfp,Hongo:2021ona,Becattini:2023ouz,Becattini:2025oyi}. 

From here, we can also derive the ``first law of spin thermodynamics''. Let us introduce the chemical potential $\mu$, the temperature $T$, the ``spin chemical potential'' $\mu_{\mu \nu}$, and the hydrodynamic flow velocity $u^\mu$ (with $u^\mu u_\mu=-1$, $u^0>0$), through the equations below:
\begin{equation}
    \begin{split}
        \alpha ={}& \mu/T \, ,\\
        \beta_{\mu} ={}& u_\mu/T \, , \\
        \Omega_{\mu \nu} ={}& \mu_{\mu \nu}/T \, .
    \end{split}
\end{equation}
Let us define the rest-frame densities of entropy, particle, energy, and spin, respectively:
\begin{equation}
    \begin{split}
        s={}& -u_\lambda s^\lambda \, , \\
        n={}& -u_\lambda N^\lambda \, , \\
        \varepsilon ={}& u_\lambda u_\mu T^{\lambda\mu} \, , \\
        \mathfrak{s}^{\mu \nu}={}& -u_\lambda \mathcal{S}^{\lambda \mu \nu} \, . \\
    \end{split}
\end{equation}
Then, contracting \eqref{covGibbsFinal} with $u_\lambda$, and setting $du_\lambda=0$, i.e., imposing that the variation does not alter the state of motion of the fluid, we finally obtain the first law of spin thermodynamics:
\begin{equation}
d\varepsilon = T ds +\mu dn + \dfrac{\mu_{\mu \nu}}{2} d\mathfrak{s}^{\mu \nu}  \, .
\end{equation}
\section{Thermodynamic analysis}\label{sec:thermodynamicanalysis}
In this section, we analyze the properties of the thermodynamic equilibrium states as they are predicted by the hydrodynamic theory. 
\subsection{Extremum principle}\label{extremumprinciple}
Consider an isolated system composed of two parts: a spinfluid and a surrounding environment. The state of thermodynamic equilibrium of the total system ``spinfluid+environment'' is the state that maximizes the total entropy, $S_{\rm tot}=S_F+S_E$ 
(the subscripts ``$F$'' and ``$E$'' refer, respectively, to spinFluid and Environment), for fixed values of all the Noether charges of the system, $Q_{\text{tot}}^I=Q^I_F+Q_E^I$. The environment is assumed to be infinitely larger than the spinfluid, and it is always in thermodynamic equilibrium with itself, meaning that there is an equation of state of the form $S_E=S_E(Q_E^I)$. 

Therefore,
\begin{equation} 
S_{\text{tot}}=S_F+S_E(Q_{\text{tot}}^J-Q^J_F) \approx S_F+S_E(Q_{\text{tot}}^J)-\dfrac{\partial S_E(Q_{\text{tot}}^J)}{\partial Q_E^I} Q_F^I \, ,
\end{equation}
where all higher order terms converge to zero in the limit of infinitely large environment (i.e., as $Q_F^I/Q_E^I \rightarrow 0$). As we can see, the requirement that $S_{\text{tot}}$ should be maximized for constant $Q_{\text{tot}}^J$ is equivalent to the requirement that the quantity $\Phi=S_F +\alpha^\star_I Q_F^I$ should be maximized for constant $\alpha^\star_I$, with
\begin{equation}
    \alpha^\star_I = -\dfrac{\partial S_E(Q_{\text{tot}}^J)}{\partial Q_E^I} \, .
\end{equation}
This may be viewed as a generalization of the minimum free energy principle \cite{Callen_book} to a system with arbitrary conserved charges\footnote{Indeed, if the only Noether charge of the system is this energy, then $\Phi=S_F-U_F/T^\star$, imply that $-T^\star \Phi$ is the Helmholtz free energy.}. In our case of interest, the functional $\Phi$ reads
\begin{equation}
\label{functional}
    \Phi= S_F +\alpha^\star N_F +b^\star_\mu P_F^\mu +\dfrac{\varpi^\star_{\mu \nu} \mathcal{J}^{\mu \nu}_F +\mathbb{A}^\star_{\mu \nu}\mathcal{S}^{\mu \nu}_F}{2} \, .
\end{equation}
In fact, the relevant Noether charges are the particle (minus antiparticle) number $N_F$, the four-momentum $P^\mu_F$, and the total angular momentum $\mathcal{J}^{\mu \nu}_F$. We have also included, among the charges, the total spin $\mathcal{S}^{\mu \nu}_F$. The latter is not a proper Noether charge. However, we are working under the assumption that the non-local collisions do not occur, so $\mathcal{S}^{\mu \nu}_F$ effectively behaves like a conserved charge (see Sec. \ref{basicassumptions}). The quantities $\alpha^\star_I = \{\alpha^\star, b^\star_\mu, \varpi^\star_{\mu \nu}/2,\mathbb{A}_{\mu \nu}^\star/2 \}$ are constants that characterize the internal state of the environment. If the constituents of the environment can experience non-local collisions, then the environment is in spin equilibrium, and we need to set $\mathbb{A}_{\mu \nu}^\star=0$.

\subsection{Equilibrium state}\label{statemech}

We can use the extremum principle introduced above to find the equilibrium states of our divergence-type theory. The procedure is standard \cite{GavassinoStabilityCarter2022,
GavassinoGENERIC2022}. First, we express $\Phi$ as a hydrodynamic integral:
\begin{equation}
    \Phi = \int_\Sigma \phi^\lambda d\Sigma_\lambda \, ,
\end{equation}
where $\Sigma$ is a spacelike surface that covers the support of the spinfluid, and the current $\phi^\lambda$ is [recall Eq.~\eqref{agmuni}]
\begin{equation}\label{philambda}
    \phi^\lambda = \chi^\lambda +(\alpha^\star-\alpha)N^\lambda + (b^\star_\mu +x^\nu \varpi^\star_{\nu \mu}-\beta_\mu)T^{\lambda \mu} + \dfrac{(\varpi^\star_{\mu \nu}+\mathbb{A}_{\mu \nu}^\star-\Omega_{\mu \nu})\mathcal{S}^{\lambda \mu \nu}}{2} \, .
\end{equation}
Then, we consider a smooth one-parameter family, $\zeta_A(\epsilon)$, of solutions of the field equations~\eqref{conservations}, where $\epsilon=0$ describes the equilibrium state. All physical quantities, including $\phi^\lambda$, can be written as functions of $\epsilon$. The constants $\alpha^\star_I$ are fixed (i.e., they do not depend on $\epsilon$). The condition that $\Phi$ should be maximized for constant $\alpha^\star_I$ implies that $\dot{\Phi}(\epsilon=0)=0$ for any choice of one parameter family defined as above (we have introduced the notation $\dot{f}=df/d\epsilon$). But $\dot{\Phi}$ is the integral of the current [recall Eq.~\eqref{diffuzzo}]
\begin{equation}\label{phidot}
    \dot{\phi}^\lambda = (\alpha^\star-\alpha)\dot{N}^\lambda + (b^\star_\mu +x^\nu \varpi^\star_{\nu \mu}-\beta_\mu)\dot{T}^{\lambda \mu} + \dfrac{(\varpi^\star_{\mu \nu}+\mathbb{A}_{\mu \nu}^\star-\Omega_{\mu \nu})\dot{\mathcal{S}}^{\lambda \mu \nu}}{2} \, .
\end{equation}
For the integral to vanish on arbitrary hypersurfaces, we need to set $\dot{\phi}^\lambda(\epsilon=0)=0$. This produces the following equilibrium conditions, which identify the equilibrium state uniquely:
\begin{equation}\label{equilibriumconditions}
    \begin{split}
        \alpha={}& \alpha^\star \, , \\
        \beta_\mu={}& b^\star_\mu +x^\nu \varpi^\star_{\nu \mu} \, , \\
        \Omega_{\mu \nu}={}& \varpi^\star_{\mu \nu}+\mathbb{A}_{\mu \nu}^\star \, . \\
    \end{split}
\end{equation}
The equilibrium conditions above generalize the zeroth law of thermodynamics \cite{GavassinoTermometri} to spinfluids. In fact, they tell us that, at equilibrium, the ``intensive'' variables $\zeta_A$ should coincide with the corresponding ``intensive'' variables of the environment. Furthermore, since $\alpha^\star_I$ are constants, Eq.~\eqref{equilibriumconditions} gives us full information about the hydrodynamic features of all equilibrium states. In fact, from \eqref{equilibriumconditions} we find that ``thermodynamic equilibrium'' implies that $\partial_\mu \alpha=0$ (i.e., diffusive equilibrium) and $\partial_{(\nu}\beta_{\mu)}=0$ (i.e., thermal equilibrium, plus absence of shear and expansion). Furthermore, if the environment is in spin equilibrium, namely if $\mathbb{A}_{\mu \nu}^\star=0$, then $\partial_{[\nu}\beta_{\mu]}=\Omega_{\nu \mu}$, which is the standard condition for spin equilibrium in fluids. These rigorous predictions of the hydrodynamic theory are in full agreement with statistical quantum field theory \cite{Speranza:2020ilk,Becattini:2023ouz,Becattini:2025oyi}. Actually, we can push the connection with statistical mechanics even further. In fact, plugging \eqref{equilibriumconditions} into \eqref{philambda}, we find that, at equilibrium, $\phi^\lambda$ equals the Massieu current $\chi^\lambda$, so that
\begin{equation}\label{staticizzo}
    \Phi(\epsilon=0) = \int_\Sigma \dfrac{\partial \chi}{\partial \beta_\lambda} d\Sigma_\lambda \, .
\end{equation}
But since $e^{\Phi(\epsilon=0)}$ coincides with the grand-canonical partition function of the fluid \cite{GibbonsHawking1977}, Eq.~\eqref{staticizzo} can be used to compute the equation of state $\chi(\zeta_A)$ from microphysics.

As a consistency check, let us prove that a field configuration $\zeta_A(x^\mu)$ that satisfies the conditions \eqref{equilibriumconditions} is necessarily a solution of the fluid equations \eqref{symmetrichyperb}. Indeed, if we expand the sum over the index $B$, Eq.~\eqref{equilibriumconditions} becomes
\begin{equation}
     M^{\lambda A \alpha}\partial_\lambda \alpha +  M^{\lambda A \mu}\partial_\lambda \beta_\mu +M^{\lambda A \mu \nu}\partial_\lambda \Omega_{\mu \nu}=0 \, .
\end{equation}
The first and the third term vanish identically, because $\partial_\lambda \alpha=0$ and $\partial_\lambda \Omega_{\mu \nu}=\partial_\lambda \varpi^\star_{\mu \nu}=0$ (since $\varpi^\star_{\mu \nu}$ is a constant). The second term also vanishes, because $\partial_{\lambda}\beta_{\mu}$ is antisymmetric in $\lambda$ and $\mu$ by the Killing condition, while $M^{\lambda A \mu}$ is symmetric, since
\begin{equation}
    M^{\lambda A \mu}= \dfrac{\partial^3 \chi}{\partial \beta_\lambda \partial \zeta_A \partial \beta_\mu} \, .
\end{equation}

\subsection{Stability of the equilibrium}

In the previous subsection, we have identified the equilibrium state as the state that makes the functional $\Phi$ stationary (for fixed $\alpha^\star_I$). However, we still need to make sure that such a stationary point is a genuine maximum. In particular, we need to verify that the quantity $E=\Phi(0)-\Phi(\epsilon)$ is positive definite. For small deviations from equilibrium, this can be done easily. In the limit of small $\epsilon$, we can express $E$ as the integral of a quadratic ``information current''~\cite{GavassinoCausality2021}\footnote{\color{black} It can be proven \cite{GavassinoCausality2021} that the quantity $E$ is the amount of microscopic information that one gains by measuring the hydrodynamic fields, which justifies the interpretation of $E^\mu$ as the related flow of information. From this perspective, the conservation equation $\partial_\mu E^\mu=0$ is just the statement that a non-dissipative theory (being reversible) conserves information.}
\begin{equation}\label{infodef}
    E^\lambda = \phi^\lambda(0)-\phi^\lambda(\epsilon) = -\dfrac{1}{2} \ddot{\phi}^\lambda(0) \epsilon^2 + \mathcal{O}(\epsilon^3) \, .
\end{equation}
To compute $\ddot{\phi}^\lambda(0)$ we only need to differentiate Eq.~\eqref{phidot} with respect to $\epsilon$ [note that equations \eqref{philambda} and \eqref{phidot} are valid for all values of $\epsilon$] and evaluate the result at $\epsilon=0$, where the equilibrium conditions \eqref{equilibriumconditions} hold.
This gives 
\begin{equation}
\ddot{\phi}^\lambda(0)= -\dot{\zeta}_A(0) \dot{N}^{\lambda A}(0)\,.    
\end{equation}
On the other hand, we know that $\dot{N}^{\lambda A}= M^{\lambda A B}\dot{\zeta}_B$. Therefore, introducing the notation $\delta \zeta_A=\zeta_A(\epsilon)-\zeta_A(0)=\dot{\zeta}_A(0)\epsilon + \mathcal{O}(\epsilon^2)$, the information current \eqref{infodef} can be expressed as follows:
\begin{equation}\label{infoquadratic}
    E^\lambda = \dfrac{1}{2} M^{\lambda A B}\,\delta \zeta_A \,\delta \zeta_B + \mathcal{O}(\epsilon^3),
\end{equation}
where $\delta \zeta_A$ can be interpreted as first-order perturbations to the fields, and $M^{\lambda AB}$ is evaluated at equilibrium. 
We would like to emphasize that we have recovered a very general fact~\cite{GavassinoCasmir2022}: the information current is a quadratic form in the perturbation fields, whose associated (symmetric) matrix $M^{\lambda AB}$ is also the matrix that defines the principal part of the field equations, see Eq.~\eqref{symmetrichyperb}. If the theory is symmetric, hyperbolic and causal, then $E^\lambda$ is future directed non-space-like [compare with Eq.~\eqref{causality}, and the discussion underneath]. But this implies that the functional $E$ is positive definite (at least for small $\epsilon$), ensuring that the equilibrium state is a genuine maximum of $\Phi$, which is what we wanted to prove.

We can use the above result to prove that the equilibrium state is dynamically stable to perturbations. In fact, in an arbitrary hydrodynamic process, the change in $\Phi$ is given by $\Delta \Phi= \Delta S_F - \alpha^\star_I \Delta Q^I_F$ (recall that $\alpha^\star_I$ are constants). If the spinfluid evolves hydrodynamically, consistently with Eq.~\eqref{conservations}, its entropy is conserved ($\Delta S_F=0$), and so are its Noether charges ($\Delta Q_F^I=0$). Hence, $\Delta \Phi=0$. As a consequence, $E=\Phi(0)-\Phi(\epsilon)$ is constant, because both $\Phi(0)$ and $\Phi(\epsilon)$ do not depend on the Cauchy surface $\Sigma$ upon which we evaluate them.\footnote{In App. \ref{conservationofinfo}, we verify explicitly that, indeed, $\partial_\lambda E^\lambda=0$ along linearized solutions of \eqref{symmetrichyperb}.} But this implies that the equilibrium state is Lyapunov stable, since we can associate with the perturbation $\delta \zeta_A$ a positive definite square integral norm that cannot grow in time. Note that this is simultaneously true in all reference frames, meaning that the spinfluid is ``covariantly stable'' \cite{GavassinoSuperluminal2021}. Furthermore, our reasoning is valid for both non-rotating and  rotating equilibrium. 
{\color{black}
\section{Generalized equation of state for a perfect spinfluid within kinetic theory}
\label{sec:Generalized equation of state}
}
{\color{black}In this section, we consider a weakly coupled dilute gas of spin-$\frac{1}{2}$ particles, modeled within kinetic theory. This assumption allows us to write down a specific expression of the generalized equation of state (generating function) by extending the Maxwell--Jüttner distribution function to the spinfluid~\cite{Florkowski:2018fap,Weickgenannt:2020aaf}: }
\begin{eqnarray}
    \chi \left(x,p,\mathfrak s\right) &=& \int \left[\underbrace{\exp \left(\beta_\mu p^\mu + \alpha  + \dfrac{\Omega_{\mu \nu}\Sigma^{\mu \nu}}{2} \right)}_{\rm particles} + \underbrace{\exp \left(\beta_\mu p^\mu -\alpha + \dfrac{\Omega_{\mu \nu}\Sigma^{\mu \nu}}{2} \right)}_{\rm antiparticles} \right] dP\, dS(p) ,\,\,\,\nonumber\\
    &=& 2\, \cosh(\alpha) \int  \exp \left(\beta_\mu p^\mu +  \dfrac{\Omega_{\mu \nu}\Sigma^{\mu \nu}}{2} \right)dP\ dS(p).
    \label{eq:EoS}
\end{eqnarray}
The phase space has been extended to include the spin degree of freedom $\mathfrak s^\mu$, in terms of which the spin dipole-moment tensor is expressed as 
\begin{equation}
 \Sigma^{\mu \nu}=\frac{1}{m}\epsilon^{\mu\nu\alpha\beta}p_\alpha \mathfrak s_\beta .  
\end{equation}
The momentum and spin integration measures are $dP = \frac{d^3 p}{(2\pi)^3 E_{\bm p}}$ and $dS(p) = \frac{m}{\pi \tilde{s}} d^4 \mathfrak s \,\delta(\mathfrak s^2-\tilde{s}^2)\delta (p \cdot \mathfrak s)$, respectively, where $E_{\bm p}=\sqrt{\bm p^2 + m^2}$, and $\tilde{s}$ is a normalization conventionally chosen as $\tilde{s} = \sqrt{\frac{1}{2}\left(1+\frac{1}{2}\right)}=\frac{\sqrt 3}{2}$ for spin-$\frac{1}{2}$ particles. The fugacity factor, $\alpha$, is the ratio of the baryon chemical potential $\mu$ over temperature $T$. We note that $f_\pm(x, p, \mathfrak s)\equiv\exp\left(\beta_\mu p^\mu \pm \alpha + \frac{\Omega_{\mu\nu}\Sigma^{\mu\nu}}{2}\right)$ is the exact local equilibrium distribution function for a fluid consisting of classical spinning tops.

The spin-hydrodynamic theory generated from \eqref{eq:EoS} is symmetric-hyperbolic and non-linearly causal, which, as described in the previous section, also implies that every equilibrium state is linearly stable. To see this, in a compact notation, we write
\begin{equation}
    \chi = \int  \left(e^{ \zeta_C p^C} + e^{\zeta_C \bar{p}^C}\right) dP\ dS(p)\,,
\end{equation}
with $p^C=\{1, p^\mu,\Sigma^{\mu \nu} \}$, $\bar{p}^C=\{-1,p^\mu,\Sigma^{\mu \nu} \}$, and $\zeta_C=\{\alpha,\beta_\mu,\Omega_{\mu\nu}/2\}$. Then
\begin{equation}
    \chi^\lambda= \int  \left(e^{ \zeta_C p^C} + e^{\zeta_C \bar{p}^C}\right) p^\lambda dP\ dS(p)\, ,
\end{equation}
and, from \eqref{agmuni}, we have
\begin{equation}
    N^{\lambda A}= \int  \left(e^{ \zeta_C p^C} p^A + e^{\zeta_C \bar{p}^C} \bar{p}^A\right) p^\lambda dP\ dS(p)\, ,
    \label{eq:currents}
\end{equation}
where $N^{\lambda A}= \{ N^\lambda, T^{\lambda \mu},\mathcal{S}^{\lambda \mu \nu} \}$.
The principal part matrix appearing in Eq.~\eqref{symmetrichyperb} is then
\begin{equation}
    M^{\lambda AB} = \int  \left(e^{ \zeta_C p^C} p^Ap^B + e^{\zeta_C \bar{p}^C} \bar{p}^A \bar{p}^B\right) p^\lambda dP\ dS(p)\, .
\end{equation}
As described in Sec.~\ref{framework}, the theory is symmetric-hyperbolic and causal when
\begin{equation}
   M^{\lambda AB}Z_A Z_B = \int  \left[ e^{ \zeta_C p^C} (p^AZ_A)^2 + e^{\zeta_C \bar{p}^C} (\bar{p}^AZ_A)^2 \right] p^\lambda dP\ dS(p)\,
   \label{eq:non-linear-causality}
\end{equation}
is future-directed and non-spacelike. But the factor in square brackets in equation~\eqref{eq:non-linear-causality} is manifestly positive definite. Therefore, $M^{\lambda AB}Z_A Z_B$ is indeed future-directed and timelike, being a continuous sum of future-directed timelike vectors.

\subsection{Evaluation of the currents}

Using the generating function \eqref{eq:EoS}, one obtains the conserved currents \eqref{agmuni} as
\begin{equation} \label{currentsKT}
    \begin{split}
    	N^\mu =& \frac{\p^2 \chi}{\p \beta_\mu \p \alpha} = 2 \sinh(\alpha) \int dP\ p^\mu e^{p^\gamma \beta_\gamma} \int dS(p)\ \exp\left( \frac{1}{2}\Omega_{\alpha\beta}\Sigma^{\alpha\beta} \right), \\
    	T^{\mu\nu} =& \frac{\p^2\chi}{\p\beta_\mu\p \beta_\nu} = 2\cosh(\alpha) \int dP\ p^\mu p^\nu e^{p^\gamma \beta_\gamma} \int dS(p)\ \exp\left( \frac{1}{2}\Omega_{\alpha\beta} \Sigma^{\alpha\beta}\right), \\
        \mathcal{S}^{\lambda\mu\nu} =& \frac{\p^2 \chi}{\p \beta_\lambda \p(\Omega_{\mu\nu}/2)} = 2 \cosh(\alpha) \int dP\ p^\lambda e^{p^\gamma \beta_\gamma}\int dS(p)\ \Sigma^{\mu\nu} \exp\left( \frac{1}{2}\Omega_{\alpha\beta} \Sigma^{\alpha\beta}\right).
    \end{split}
\end{equation}
We note that, to first-order and to second-order in $\Omega_{\mu\nu}$, these currents reproduce the results mentioned in~\cite{Florkowski:2019qdp,Singh:2020rht,Florkowski:2021wvk,Weickgenannt:2022zxs,Florkowski:2024bfw}. 

However, the expressions \eqref{currentsKT} can be evaluated numerically to all orders in $\Omega_{\mu\nu}$, and, as mentioned above, are applicable to a fluid of classical spinning tops in the absence of spin-orbit coupling (due to the symmetric nature of $T^{\mu\nu}$). It is worth noting here that the integrals can fail to converge when components of $\Omega_{\mu\nu}$ are sufficiently large, because the exponent $\frac{1}{2}\Omega_{\alpha\beta}\Sigma^{\alpha\beta}$ can be positive, as described below.

With the goal of numerically evaluating the full currents \eqref{currentsKT}, we first perform the spin integrals appearing in these expressions. Let us consider the integral
\begin{align}
    F_0(p; \Omega) \equiv \frac{1}{2}\int dS(p)\ \exp\left( \frac{1}{2} \Omega_{\alpha\beta} \Sigma^{\alpha\beta}\right).
\end{align}
We introduce the polarization four-vector
\begin{align}
    P^\mu = -\frac{1}{2m}\epsilon^{\mu\alpha\beta\gamma} p_\alpha \Omega_{\beta\gamma},
\end{align}
which, in the particle rest frame (PRF) defined by $p^\mu = (m, \bm 0)$, reproduces the components of the polarization three-vector \cite{Florkowski:2018fap}. In terms of $P^\mu$, the exponent $\frac{1}{2}\Omega_{\alpha\beta}\Sigma^{\alpha\beta}$ can be expressed as $P_\mu \mathfrak s^\mu$. Furthermore, by construction, the temporal component of $P^\mu$ vanishes in the PRF, so $P_\mu \mathfrak s^\mu = \bm P \cdot  \mathfrak s$ in the PRF. We evaluate
\begin{equation} \label{F0}
    \begin{split}
    F_0(p; \Omega) &= \frac{1}{2}\int dS(p)\ e^{\bm P \cdot \mathfrak  s} \\
    &= \frac{m}{2\pi\tilde{s}}\int d^3 \mathfrak s\ e^{\bm P \cdot \mathfrak s}\int d \mathfrak s^0\ \delta(\mathfrak s^2-\tilde{s}^2)\delta(-m \mathfrak s^0) \\
    &= \frac{1}{2\pi \tilde{s}} \int d^3 \mathfrak s\ e^{\bm P \cdot \mathfrak s}\delta(|\mathfrak s|^2-\tilde{s}^2) \\
    &= \frac{\sinh(\tilde{s} P)}{\tilde{s} P}\,,
    \end{split}
\end{equation}
which was also obtained in~\cite{Florkowski:2018fap}.
In writing the last line, we have returned to an arbitrary Lorentz frame and defined $P= \sqrt{P^\mu P_\mu}$.

Next, we consider the integral
\begin{align}
    G^{\mu\nu}(p; \Omega) \equiv \frac{1}{2}\int dS(p)\  \Sigma^{\mu\nu} \exp\left( \frac{1}{2}\Omega_{\alpha\beta}\Sigma^{\alpha\beta}\right).
\end{align}
This can be evaluated by noticing that $G^{\mu\nu} = \frac{\p F_0}{\p(\Omega_{\mu\nu}/2)} = \frac{\p F_0}{\p P}\frac{\p P}{\p P^\lambda} \frac{\p P^\lambda}{\p(\Omega_{\mu\nu}/2)}$. The result is
\begin{align}
    G^{\mu\nu}(p; \Omega) &= \frac{1}{m^2}\left( m^2\Omega^{\mu\nu} - 2p^\alpha p^{[\mu}\Omega^{\nu]}{}_\alpha\right)F_1(p; \Omega),
\end{align}
where
\begin{align} \label{F1}
    F_1(p; \Omega) &= \frac{1}{P}\frac{\p F_0}{\p P} = \frac{\tilde{s} P\cosh(\tilde{s} P)-\sinh(\tilde{s} P)}{\tilde{s} P^3}\,.
\end{align}
For use in the next section, we also evaluate
\begin{align}
    H^{\mu\nu}_{\rho\sigma}(p; \Omega) &\equiv \frac{1}{2}\int dS(p) \Sigma^{\mu\nu} \Sigma_{\rho\sigma} \exp\left(\frac{1}{2}\Omega_{\alpha\beta}\Sigma^{\alpha\beta}\right).
\end{align}
Manipulations similar to the above lead to
\begin{align}
     H^{\mu\nu}_{\rho\sigma}(p; \Omega) &= \frac{1}{m^2}\left[ - F_1p_\alpha p^\gamma\delta^{\mu\nu\alpha}_{\rho\sigma\gamma} + \frac{F_2}{m^2}\left(m^2\Omega^{\mu\nu}-2p^\alpha p^{[\mu}\Omega^{\nu]}{}_\alpha \right)\left(m^2\Omega_{\rho\sigma}-2p_\alpha p_{[\rho}\Omega_{\sigma]}{}^\alpha \right) \right],
\end{align}
where $\delta^{\mu\nu\alpha}_{\rho\sigma\gamma}$ is the generalized Kronecker delta, equal to zero unless $\mu\nu\alpha$ is a permutation of $\rho\sigma\gamma$ (in which case it equals the parity of the permutation), and
\begin{align} \label{F2}
    F_2(p; \Omega) &= \frac{1}{P}\frac{\p F_1}{\p P} = \frac{(3+P^2 \tilde{s}^2 ) \sinh(\tilde{s} P) - 3 \tilde{s} P \cosh(\tilde{s} P)}{\tilde{s} P^5}.
\end{align}
With the spin integrals evaluated, every component of each of the currents \eqref{currentsKT} is proportional to a momentum integral of the form
\begin{align} \label{integralsoftheform}
    \int dP\ p^{\alpha_1}\cdots p^{\alpha_n}e^{p^\gamma \beta_\gamma}F_a(p; \Omega), \qquad a=0,1\,.
\end{align}
To numerically evaluate such an integral at a given spacetime point, it is convenient to go to the local rest frame (LRF) of the fluid, i.e., the Lorentz frame in which $\beta^\mu=(\beta, \bm 0)$. Then the spin potential, being an antisymmetric tensor, can be decomposed in terms of two spatial vectors $\bm e$ and $\bm b$ in the LRF:
\begin{align} \label{eandb}
    \Omega_{\mu\nu} = \begin{pmatrix}
    0 & e_x & e_y & e_z \\
    -e_x & 0 & b_z & -b_y \\
    -e_y & - b_z & 0 & b_x \\
    -e_z & b_y & -b_x & 0
    \end{pmatrix}.
\end{align}
By spatial rotation of the coordinate axes, $\bm b$ can always be brought into alignment with the $z$-axis. Furthermore, we specialize for the remainder of the article to the case in which $\bm e$ is parallel to $\bm b$ (or that it is very small). Doing so is not necessary, but it allows \eqref{integralsoftheform} to be reduced to a one-dimensional integral, as will be seen shortly. That is, returning to the original notation, we take
\begin{align}
    \Omega_{\mu\nu} = \begin{pmatrix}
    0 & 0 & 0 & \Omega_{03} \\
    0 & 0 & \Omega_{12} & 0 \\
    0 & - \Omega_{12} & 0 & 0 \\
    -\Omega_{03} & 0 & 0 & 0
    \end{pmatrix}
\end{align}
in the LRF.

It is then straightforward to show that the $p^\mu$-dependence of $F_i$ is only via ${p_\perp = \sqrt{p_x^2 + p_y^2}}$; in particular, we have
\begin{align}
    P = \frac{1}{m}\sqrt{m^2 (\Omega_{12})^2 + \left[ (\Omega_{12})^2 + (\Omega_{03})^2\right]p_\perp^2}.
\end{align}
Factors of $p_x$ in the integrand of \eqref{integralsoftheform} must then come in pairs, or else the integral will vanish by symmetry, and factors of $p_x^2$ can be replaced by $p_\perp^2/2$. The same is true for $p_y$. With this in mind, we redefine \eqref{integralsoftheform} by the dimensionless integral
\begin{align} \label{Idef}
    I_a^{\ell mn}&\equiv \beta^{2+\ell+2(m+n)}\int dP\ (E_{\bm p})^\ell (p_\perp)^{2m} (p_z)^{2n} e^{-\beta E_{\bm p}} F_a(p_\perp; \Omega_{12}, \Omega_{03}),
\end{align}
making a slight notational change in the arguments of $F_a$ to reflect their dependence on only $p_\perp$, $\Omega_{12}$, and $\Omega_{03}$. The $p_z$-integral can be evaluated first, followed by the (trivial) integral over the angular coordinate in the $p_x$-$p_y$ plane, followed finally by the integral over the radial coordinate $p_\perp$. The end result is
\begin{align} \label{I}
    I_a^{\ell mn} &= (-1)^\ell \frac{(2n)!}{\pi^22^{n-1} n!}\int_z^\infty dx\ x^{1+\ell + 2n}(x^2-z^2)^m \mathcal K_{\ell n}(x) \tilde F_a(x; \Omega_{12}, \Omega_{03}),
\end{align}
where $z$ is the dimensionless quantity defined as the product of mass $m$ and inverse temperature $\beta$, $z=m\,\beta = m/T$. This quantity can be used as a controlling parameter to go from the non-relativistic (low-temperature) limit to the relativistic (high-temperature) limit. We also define $\mathcal K_{\ell n}=\frac{\p^\ell}{\p x^\ell}\left[ x^{-n}K_n(x)\right]$ (with $K_n(x)$ being the modified Bessel function of the second kind), and $\tilde F_a(x; \Omega_{12}, \Omega_{03})= F_a(p_\perp=\sqrt{x^2-z^2}/\beta; \Omega_{12}, \Omega_{03})$.

In this way, by numerically evaluating \eqref{I} for various values of $\ell$, $m$, and $n$, we can obtain all components of the currents \eqref{currentsKT}. However, at large $x$, $\mathcal K_{\ell n}$ goes like $x^{-n-1/2} e^{-x}$ while $\tilde F_a$ goes like $x^{-2a-1}\exp\left(x\, \tilde{s}\sqrt{(\Omega_{12})^2+(\Omega_{03})^2}/z\right)$, so the currents \eqref{currentsKT} diverge as $z$ approaches $\tilde{s} \sqrt{(\Omega_{12})^2+(\Omega_{03})^2}$ from above\footnote{The divergence of the integral here is analogous to the divergence of the partition function in a Bose gas when we try to push the chemical potential above the energy of the one-body ground state. Physically, this divergence is related to the fact that it takes an infinite density of energy or particles to reach these states, so a real fluid will never evolve to that point.}. In the case of general $\Omega_{\mu\nu}$, the condition for convergence becomes~\footnote{See Refs.~\cite{Bhadury:2025boe,Drogosz:2025ihp} for related work on the range of applicability of perfect spin hydrodynamics.}
\begin{align} \label{convergence}
    z > \tilde{s} \sqrt{\bm e^2+ \bm b^2},
\end{align}
where $\bm e$ and $\bm b$ are defined in \eqref{eandb}, and in our special case, where we work in the rest frame of fluid, we have $\bm e = (0, 0, 0, \Omega_{03})$ and $\bm b = (0, 0, 0, \Omega_{12})$.
\\

\begin{figure}[h]
    \centering
    \includegraphics[width=0.9\textwidth]{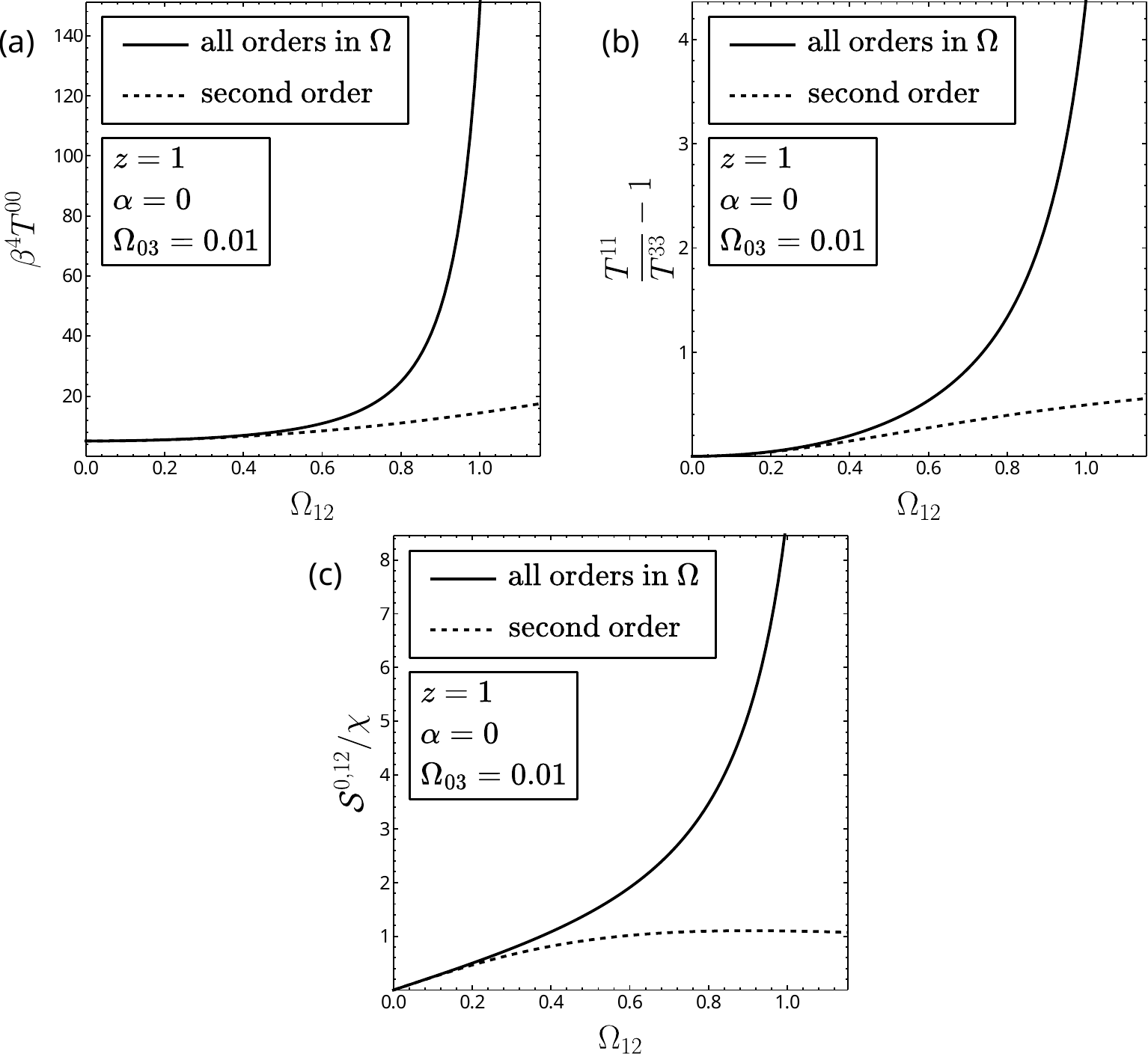}
    \caption{\textbf{Energy density, pressure anisotropy, and spin per particle.} In all plots, $z=1$, $\alpha=0$, $\Omega_{03}=0.01$, and the maximum value on the horizontal axis, $\Omega_{12}\approx \tilde{s}^{-1}=2/\sqrt{3}$, saturates the bound \eqref{convergence}. The solid curves are numerically exact to all orders in $\Omega_{\mu\nu}$, while the dashed lines are obtained from second-order-in-$\Omega_{\mu\nu}$ truncations. \textbf{(a)} Energy density, \textbf{(b)} pressure anisotropy, and \textbf{(c)} $z$-component of spin per particle are shown.}
    \label{fig:currents}
\end{figure}

In Fig.~\ref{fig:currents}, we show the energy density component $T^{00}$ (normalized by temperature), the pressure anisotropy $(T^{11}-T^{33})/T^{33}$, and the $z$-component of spin per particle $\mathcal S^{0, 12}/\chi$. In the latter, $\chi$ is the generating function, which is, physically, the density of all particles counting both particles and antiparticles with the same sign, as can be understood directly from Eq.~\eqref{eq:EoS}. For the numerical analysis, we kept $z=1$, $\alpha=0$ with fixed $\Omega_{03}=0.01$. The maximum value on the horizontal axis, $\Omega_{12}\approx \tilde{s}^{-1}=2/\sqrt{3}$, saturates the bound \eqref{convergence}.

We observe that the spin per particle increases with $\Omega_{12}$, and so do the energy density and pressure anisotropy. From Fig.~\ref{fig:currents}(b), we note that the pressure in the $x$-$y$ plane exceeds the pressure in the $z$-direction when the spin polarization is large. The pressure anisotropy appears due to the spin contribution to the energy-momentum tensor, which vanishes if we truncate to first order in $\Omega_{\mu\nu}$~\cite{Florkowski:2019qdp}. Hence, the pressure anisotropy vanishes for $\Omega_{12}=0$.

In Fig.~\ref{fig:currents}(c), the spin per particle is unbounded. While in the PRF the $z$-component of a particle's spin angular momentum is bounded by $\tilde{s}$, it can be larger in the LRF of the fluid. For example, one can show directly that, in the LRF, the $z$-component of the spin of a particle traveling in the $z$-direction with speed $v$ is $\Sigma^{12} = \Sigma^{12}_\text{PRF} \frac{1+v^2}{1-v^2} > \Sigma^{12}_\text{PRF}$, where $\Sigma^{12}_\text{PRF}$ is the $z$-component of spin angular momentum evaluated in the PRF. In the non-relativistic limit $z \gg 1$ and $|T^{00}/(m\chi) - 1| \ll 1$, we find from \eqref{currentsKT} that the spin per particle is bounded from above by $\tilde{s} = \frac{\sqrt 3}{2}$, as required.

While we have focused for simplicity on the case in which $\bm e$ is parallel to $\bm b$, this calculation can in principle be repeated as a function of the angle between $\bm e$ and $\bm b$ by evaluating multi-dimensional integrals. Such a calculation is a prerequisite for solving the equations of motion~\eqref{symmetrichyperb}.

\subsection{Linearization around a non-trivial equilibrium state}

To demonstrate the causality and stability of this theory to all orders in $\Omega_{\mu\nu}$, we study its equations of motion linearized around a non-trivial equilibrium state. Specifically, we specialize to a non-rotating fluid in the global rest frame, i.e.~$\beta^\mu = (\beta, \bm 0)$, and we take both the fugacity $\alpha$ and spin potential $\Omega_{\mu\nu}$ to be non-zero and homogeneous in the equilibrium state. For simplicity, as in the previous section, we also assume that the only non-zero components of $\Omega_{\mu\nu}$ are $\Omega_{03}$ and $\Omega_{12}$, up to antisymmetry. As described in Sec. \ref{statemech}, such a state is an equilibrium state in the absence of spin-orbit coupling due to the separate conservation of spin and orbital angular momenta.

The linearized equations of motion follow directly from \eqref{symmetrichyperb}:
\begin{align} \label{lineom}
    M^{\mu,AB}\p_\mu \delta \zeta_B = 0,
\end{align}
where $\delta\zeta_B = (\delta \alpha, \delta \beta_\lambda, \delta\Omega_{\mu\nu}/2)$ and
\begin{align} \label{MKT}
	M^{\mu, AB} =
	\begin{pmatrix}
		\coth(\alpha)N^\mu & \tanh(\alpha)T^{\mu\lambda} & \tanh(\alpha) \mathcal S^{\mu,\alpha\beta} \\
		\tanh(\alpha)T^{\mu\nu} & \frac{\p T^{\mu\nu}}{\p \beta_\lambda} & \frac{\p \mathcal S^{\mu,\alpha\beta}}{\p \beta_\nu} \\
		\tanh(\alpha)\mathcal S^{\mu, \alpha\beta} & \frac{\partial \mathcal S^{\mu,\alpha\beta}}{\p \beta_\lambda} & \frac{\p\mathcal S^{\mu,\alpha\beta}}{\p (\Omega_{\rho\sigma}/2)}
	\end{pmatrix}.
\end{align}
Here, and for the remainder of this section, $\alpha$, $\beta_\lambda$, and $\Omega_{\mu\nu}$ denote global-equilibrium quantities, while $\delta\alpha$, $\delta\beta_\lambda$ and $\delta \Omega_{\mu\nu}$ denote small perturbations around the former.

The components of $M^{\mu,AB}$ can be expressed in terms of the one-dimensional integrals $I_a^{\ell mn}$ defined in \eqref{Idef}. We solve the linearized equations of motion \eqref{lineom} with the Fourier ansatz
\begin{align} \label{fourier}
    \delta \zeta_B = \delta \zeta_B^{(0)}e^{i(\bm k \cdot \bm r - \omega t)}\,,
\end{align}
to obtain the dispersion relations $\omega_q(\bm k)$, where $q=1, \dots, 11$ indexes the modes and $\bm k$ is the wave-vector. The 11 modes reflect the $1+4+6$ degrees of freedom of the theory, corresponding respectively to the scalar $\alpha$, the four components of $\beta^\mu$, and the six independent components of $\Omega^{\mu\nu}$. The $\omega_q(\bm k)$ are linear in $k=|\bm k|$ and depend on the direction of $\bm k$ only through the polar angle $\theta_{\bm k} = \arccos(k_z/k)$. Therefore, the dispersion relations can be written as
\begin{align}
    \omega_q(\bm k) = v_q(\theta_{\bm k})k,
\end{align}
where $v_q(\theta_{\bm k})$ is the $q^\text{th}$ propagation speed.
\begin{figure}[h]
    \centering
    \includegraphics[width=0.9\textwidth]{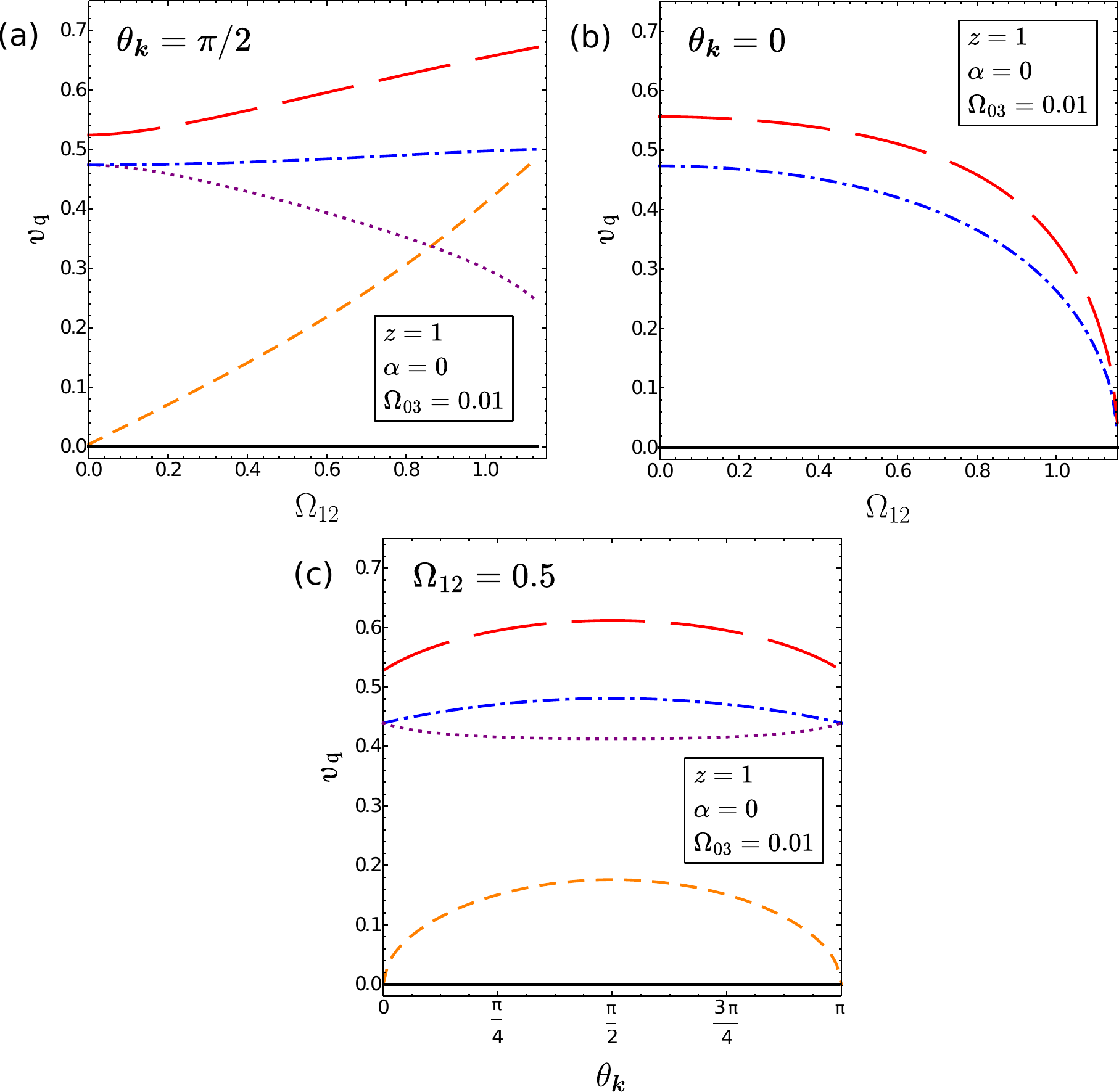}
    \caption{\label{fig:speeds} \textbf{Propagation speeds.} Parameter values are $z=1$, $\alpha=0$, and $\Omega_{03}=0.01$, as in Fig.~\ref{fig:currents}. \textbf{(a-b)} Propagation speeds $v_q$ as functions of spin potential $\Omega_{12}$ when \textbf{(a)} $\theta_{\bm k}=\pi/2$ and \textbf{(b)} $\theta_{\bm k} = 0$. In \textbf{(a)}, the black (solid) curve is triply degenerate, i.e., there are three non-propagating modes. Not pictured are the four modes obtained from the red (long-dashed), blue (dot-dashed), purple (dotted), and orange (short-dashed) curves by flipping the sign of $v_q$, which are present due to time-reversal symmetry and result in a total of $4+3+4=11$ independent modes, as described in the text. When the direction of the wave-vector is tuned from $\theta_{\bm k}=\pi/2$ to its value $\theta_{\bm k}=0$ in \textbf{(b)}, the mode corresponding to the orange (short-dashed) curve becomes non-propagating, and the modes corresponding to the blue (dot-dashed) and purple (dotted) curves become degenerate. \textbf{(c)} Propagation speeds as a function of propagation direction $\theta_{\bm k}$ when $\Omega_{12} = 0.5$, demonstrating degeneracies at $\theta_{\bm k} = 0$ and $\theta_{\bm k}=\pi$.}
\end{figure}

The propagation speeds $v_q(\theta_{\bm k})$ are plotted in Fig.~\ref{fig:speeds} for varying $\theta_{\bm k}$. All dispersion relations $\omega_q(\bm k)$ are purely real, reflecting the time-reversal symmetry of ideal spin hydrodynamics. In particular, since every mode $q$ satisfies $\operatorname{Im}[\omega_q(\bm k)]\le 0$ for all $\bm k$, the solutions \eqref{fourier} remain bounded for all time, i.e., the hydrodynamic stability of the equilibrium state under consideration is ensured. Furthermore, we observe that $v_q\le 1$ for all modes $q$, meaning that linear perturbations respect causality. It is important to note that these conclusions are demonstrated to all orders in the equilibrium spin potential $\Omega^{\mu\nu}$. While here we have only demonstrated linear hydrodynamic stability and linear causality around a restricted class of equilibrium states, we emphasize that the symmetric hyperbolicity of the full equations of motion \eqref{symmetrichyperb} guarantees linear thermodynamic stability (in the sense of Sec. \ref{statemech}) around all equilibrium states and guarantees causality in the non-linear regime.

We now consider the structure of the modes whose dispersion relations are depicted in Fig.~\ref{fig:speeds}. The parameter values are $z=1$, $\alpha=0$, and $\Omega_{03}=0.01$, as in Fig.~\ref{fig:currents}.
Let's first focus on Fig.~\ref{fig:speeds}(a) where we plot the propagation speeds $v_q$ as functions of spin potential $\Omega_{12}$ with $\theta_k=\pi/2$ i.e., the direction of propagation $\bm k/k$ is perpendicular to the shared direction of $\bm e$ and $\bm b$, in the notation of \eqref{eandb}. That is, $\bm k$ lies in the $x$-$y$ plane; without loss of generality, we take $\bm k$ to lie along the $x$-axis. One can associate the modes with spin waves, sound waves, shear waves, and charge-density waves by examining their mode vectors $\delta \zeta_{B}^{(0)}$ in the limit of vanishing spin potential, $\Omega_{12}, \Omega_{03}\rightarrow 0$.
We have three degenerate non-propagating modes $(v_q=0)$ represented by black (solid) curve and four propagating modes represented by red (long-dashed), blue (dot-dashed), purple (dotted), and orange (short-dashed) curves.
Not pictured are the other four propagating modes obtained from the red (long-dashed), blue (dot-dashed), purple (dotted), and orange (short-dashed) curves by flipping the sign of $v_q$, which are present due to time-reversal symmetry and result in a total of $3+4+4=11$ independent modes.

In the limit $(\Omega_{12}, \Omega_{03}\rightarrow 0)$, we observe from Fig.~\ref{fig:speeds}(a) that the blue (dot-dashed) and purple (dotted) modes become degenerate, i.e., they have the same speed of propagation, which we denote by $v_\text{sw}$. These are propagating spin waves: at vanishing spin potential the blue (dot-dashed) mode has $\delta \Omega_{13} = -2v_\text{sw}\delta \Omega_{03}$ with all other perturbations (up to antisymmetry) equal to zero, and the purple mode has $\delta \Omega_{12}=2\kappa v_\text{sw} \delta \Omega_{02}$ as its only nonzero components. Their shared propagation speed $v_\text{sw}$ recovers the result of Ref.~\cite{Ambrus:2022yzz}. As $\Omega_{12}$ increases, the blue (dot-dashed) mode acquires components in $\delta \alpha$, $\delta \beta^0$, and $\delta \beta^1$, i.e., it acquires oscillations in fugacity, temperature, and longitudinal flow velocity. On the other hand, the purple (dotted) mode continues to have only $\delta \Omega_{12}$ and $\delta \Omega_{02}$ nonzero, albeit in a different ratio.

There are also three \emph{non-propagating} spin modes $(v_q=0)$ at vanishing spin potential, as previously identified in Ref.~\cite{Ambrus:2022yzz}. In particular, the orange (short-dashed) mode involves only $\delta \Omega_{23}$, and upon turning on $\Omega_{12}$ becomes propagating and acquires a component in $\delta \beta^3$. That is, it becomes a hybrid spin-shear wave. The remaining shear wave at vanishing spin potential, involving only $\delta \beta^2$, mixes with the non-propagating charge-density wave and remains non-propagating when $\Omega_{12}$ is increased from zero. Similarly, the red (long-dashed) mode evolves from the sound wave at $\Omega_{12}=0$.

Finally, we note that it is possible to obtain simple asymptotic expressions for the propagation speeds $v_q(\theta_{\bm k}=\pi/2)$ in the limit $z\rightarrow \infty$ with $\Omega_{03}=0$ and keeping $\Omega_{12}\neq 0$ fixed, extending results found in Ref.~\cite{Ambrus:2022yzz}. The blue (dot-dashed) mode has speed $v\rightarrow 1/\sqrt{2z}$, the purple (dotted) mode has $v\rightarrow \frac{1}{\sqrt z}\sqrt{\frac{(\Omega_{12})^2(\hat F_2 \hat F_0-(\hat F_1)^2) + \hat F_0 \hat F_1}{(\Omega_{12})^2(\hat F_2 \hat F_0-(\hat F_1)^2) + 2\hat F_0 \hat F_1}}$, the orange (short-dashed) mode has $v\rightarrow 1/\sqrt{z}$, and the red (long-dashed) mode has $v\rightarrow \frac{|\Omega_{12}|}{z^{3/2}}\sqrt{\frac{\hat F_1}{\hat F_0}}$. In these expressions, $\hat F_a$ denotes equations \eqref{F0}, \eqref{F1}, and \eqref{F2} with $P$ replaced by $\Omega_{12}$.

When the direction of the wave-vector is $\theta_{\bm k}=0$, see Fig.~\ref{fig:speeds}(b), the mode corresponding to the orange (short-dashed) curve becomes non-propagating, and the modes corresponding to the blue (dot-dashed) and purple (dotted) curves become degenerate. For large values of $\Omega_{12}$, all modes become degenerate and non-propagating $(v_q=0)$.
In Fig.~\ref{fig:speeds}(c), we show how propagation speeds behave as a function of propagation direction $\theta_{\bm k}$ when $\Omega_{12} = 0.5$, demonstrating degeneracies of the modes at $\theta_{\bm k} = 0$ and $\theta_{\bm k}=\pi$, confirming our analysis for Figs.~\ref{fig:speeds}(a) and \ref{fig:speeds}(b).
\section{Conclusion}
\label{sec:conclude}
In this work, we have developed a novel framework for describing spin-polarized relativistic fluids by formulating a divergence-type theory that systematically incorporates spin degrees of freedom. The key advantage of our approach is that the resulting equations of motion possess a symmetric hyperbolic structure, guaranteeing nonlinear causality and the local well-posedness of the initial value problem. These mathematical properties are crucial for ensuring the predictivity and stability of the theory. Therefore, our framework can be used for numerical analyses and phenomenological applications. 

Another important aspect of our work lies in the structure of the constitutive relations and equations of motion descending from the generalized Maxwell-J\"uttner generating function \eqref{eq:EoS}. In particular, the theory remains non-linearly causal and symmetric-hyperbolic to all orders of the spin potential, as long as the constraint in Eq. \eqref{convergence} among the thermodynamic and spin variables is respected. We expect that this constraint will not be violated in the context of heavy-ion collisions \cite{Becattini:2020ngo,Das:2022lqh,Arslandok:2023utm,Becattini:2024uha,Niida:2024ntm}.

Our construction thus provides a rigorous framework for investigating a wide range of phenomena related to spin dynamics in relativistic fluids. It opens the way for future studies of spin polarization effects in high-energy nuclear collisions and other systems where relativistic spin hydrodynamics may play a significant role. {\color{black}Additionally, we expect that the framework laid out here can be naturally extended to incorporate dissipative effects and dynamical electromagnetic fields \cite{Israel:1978up, Singh:2022ltu, Bhadury:2022ulr}, offering promising directions for further theoretical development.}

\begin{acknowledgments}
We acknowledge Jorge Noronha and the participants of the Galileo Galilei Institute workshop ``Foundations and Applications of Relativistic Hydrodynamics''  for fruitful discussions. We thank the Galileo Galilei Institute for Theoretical Physics for the hospitality and the INFN for partial support during the completion of this work. N.A. is partly supported by the U.S. Department of Energy, Office of Science, Office for Nuclear Physics under Award No.~DE-SC0023861. L.G. is partially supported by a Vanderbilt Seeding Success Grant. R.S. acknowledges the kind hospitality and support of the Institute for Theoretical Physics, Goethe University Frankfurt, and CERN theory department where part of this work is completed and is supported partly by a postdoctoral fellowship of West University of Timișoara, Romania. E.S. was supported by the Rita Levi Montalcini program of the Italian Ministry of University and Research. E.S. has received funding from the European Union’s Horizon Europe research and innovation program under the Marie Sk\l odowska-Curie grant agreement No. 101109747. 
\end{acknowledgments}

\appendix

\section{Constrained derivatives}\label{AAA}

Consider a function $h=h(\Omega_{\mu \nu})$, where $\Omega_{\mu \nu}$ is an antisymmetric $4\times 4$ tensor. What does the notation
\begin{equation}\label{partiuzzi}
\dfrac{\partial h}{\partial \Omega_{12}}
\end{equation}
stand for? The answer is not obvious, because $\Omega_{21}$ coincides with $-\Omega_{12}$, and it is not clear whether we should hold $\Omega_{21}$ constant during partial differentiation, or whether we should change it together with $\Omega_{12}$. Here, we explain how one should interpret the symbol in Eq.~\eqref{partiuzzi} to make \eqref{agmuni} consistent with the divergence-type formalism.

Our starting point is the differential $dh=h^{\mu \nu}d\Omega_{\mu \nu}$. This equation should be interpreted as follows: The exterior derivative of the function $h$ is a linear combination of the exterior derivatives of the $16$ functions $\Omega_{\mu \nu}$. An expression of this kind can always be written, because $h$ is a function of $\Omega_{\mu \nu}$. However, such an expression is not unique, because $d\Omega_{\mu \nu}$ (regarded as covectors) are not linearly independent, since $d\Omega_{\mu \nu}=-d\Omega_{\nu \mu}$. This implies that there are infinite possible choices of $h^{\mu \nu}$ that are consistent with the formal expression $dh=h^{\mu \nu}d\Omega_{\mu \nu}$. On the other hand, one and only one is antisymmetric. Therefore, we \emph{define} the partial derivative of $h$ with respect to $\Omega_{\mu \nu}$ as that unique antisymmetric choice of $h^{\mu \nu}$. This is coherent with the fact that, in equations \eqref{covariantGibbs} and \eqref{diffuzzo}, the conjugate variables $\Omega_{\mu \nu}/2$ and $\mathcal{S}^{\lambda \mu \nu}$ are \emph{both} antisymmetric in $\mu$ and $\nu$. Note that, since $h^{\mu \nu}$ and $d\Omega_{\mu \nu}$ are both antisymmetric, the terms, say, $h^{12}d\Omega_{12}$ and $h^{21}d\Omega_{21}$ coincide. Hence, when we write the expression $h=h^{\mu \nu}d\Omega_{\mu \nu}$ explicitly, we get
\begin{equation}
    dh = 2h^{01}d\Omega_{01} + 2h^{02}d\Omega_{02}+2h^{03}d\Omega_{03} + 2h^{12}d\Omega_{12}+2h^{13}d\Omega_{13} + 2h^{23}d\Omega_{23}\,.
\end{equation}
For this reason, one defines the spin potential in such a way to have a factor $1/2$ in the last term of Eq.~\eqref{covGibbsFinal}: this cancels the factors $2$ that come from the antisymmetry of the spin current. 

\section{Conservation of information}\label{conservationofinfo}

The matrices $M^{\lambda AB}$ are just the third derivatives of $\chi$ with respect to $(\beta_\lambda,\zeta_A, \zeta_B)$.  Hence, if we linearize Eq.~\eqref{symmetrichyperb} around equilibrium, and we take the divergence of Eq.~\eqref{infoquadratic}, we obtain the two equations below:
\begin{equation}
    \begin{split}
\dfrac{\partial^3 \chi}{\partial \beta_\lambda \partial \zeta_A \partial \zeta_B} \partial_\lambda \delta \zeta_B + \dfrac{\partial^4 \chi}{\partial \beta_\lambda \partial \zeta_A \partial \zeta_B \partial \zeta_C} \delta \zeta_C \partial_\lambda \zeta_B &= 0 \, , \\
\dfrac{\partial^3 \chi}{\partial \beta_\lambda \partial \zeta_A \partial \zeta_B} \delta \zeta_A \partial_\lambda \delta \zeta_B + \dfrac{1}{2} \dfrac{\partial^4 \chi}{\partial \beta_\lambda \partial \zeta_A \partial \zeta_B \partial \zeta_C} \delta \zeta_A \delta \zeta_B \partial_\lambda \zeta_C &= \partial_\lambda E^\lambda  \, .  \\
    \end{split}
\end{equation}
On the other hand, at equilibrium we have that $\partial_\lambda \zeta_B=\{0,\partial_\lambda \beta_\mu,0 \}$. Furthermore, since $\partial_\lambda \beta_\mu$ is antisymmetric  under the exchange between $\lambda$ and $\mu$, when we contract $\partial_\lambda \beta_\mu$ with the fourth derivative of $\chi$ (which is symmetric), the result is zero. Therefore, the equations above simplify to
\begin{equation}
    \begin{split}
\dfrac{\partial^3 \chi}{\partial \beta_\lambda \partial \zeta_A \partial \zeta_B} \partial_\lambda \delta \zeta_B  &= 0 \, , \\
\dfrac{\partial^3 \chi}{\partial \beta_\lambda \partial \zeta_A \partial \zeta_B} \delta \zeta_A \partial_\lambda \delta \zeta_B &= \partial_\lambda E^\lambda  \, .  \\
    \end{split}
\end{equation}
Contracting the first equation with $\delta \zeta_A$ immediately results in $\partial_\lambda E^\lambda =0$.
\bibliographystyle{utphys}
\bibliography{Biblio.bib}
\end{document}